\documentclass[11pt]{article}
\usepackage[T1]{fontenc}
\usepackage[utf8]{inputenc}
\usepackage{lmodern}
\usepackage{geometry}
\usepackage{amsmath,amssymb,amsthm}
\usepackage{graphicx,booktabs,tabularx,array,multirow}
\usepackage{algorithm,algpseudocode}
\usepackage[numbers,sort&compress]{natbib}
\usepackage{xcolor}
\usepackage{xurl}
\usepackage[unicode,hidelinks]{hyperref}
\usepackage{url}
\newcolumntype{Y}{>{\raggedright\arraybackslash}X}
\newtheorem{definition}{Definition}
\newtheorem{lemma}{Lemma}
\newtheorem{theorem}{Theorem}
\newtheorem{proposition}{Proposition}
\algrenewcommand\algorithmicrequire{\textbf{Require:}}
\algrenewcommand\algorithmicensure{\textbf{Ensure:}}
\algrenewcommand{\algorithmiccomment}[1]{\hfill{\normalfont\itshape$\triangleright$~#1}}
\algblockdefx[Try]{Try}{EndTry}{\textbf{try}}{\textbf{end try}}
\algcblockdefx[Try]{Try}{Catch}{EndTry}[1]{\textbf{catch} #1}{\textbf{end try}}
\newcommand{\artifacturl}{\url{https://github.com/vectortc/Agentic-Security-Regulated-Finance}}
\title{Agent Security Meets Regulatory Reality: A Practitioner Systematization of Autonomous-Agent Threats and Controls in Regulated Financial Systems}
\author{Krishna Mohan Guda Nagavenkata Srinivasa\\Satya Phanindra Kalaga\\Karthik Guda Vishnu\\[4pt]\small VectorTech Consulting Inc., 710 Lakeway Drive, Suite 200, Sunnyvale, CA 94085, USA\\\small Correspondence: krishnamohanguda@gmail.com}
\date{}
\begin{document}
\maketitle
\begin{abstract}
Large language model agents are entering regulated financial systems, yet research on their attack surface remains predominantly laboratory-based, and practitioner guidance is neither peer-reviewed nor connected to a formal threat model. Using production evidence, we map six agentic threat categories, namely prompt injection, identity and authorization, action auditability, tool abuse, data residency, and boundary policy enforcement, to the specific control obligations imposed by US and EU financial regulation (ECOA and Regulation B, EU AI Act, GDPR Article 22, and FINRA’s 2026 agent guidance), showing how legal accountability increases their operational significance. We distill these obligations into five runtime control invariants with reference algorithms that enforce them at auditable architectural boundaries. We develop a formal model expressing each invariant as a predicate over a case-bound event chain and prove the audit-reconstruction gate is sound and complete with respect to these invariants: compliance is decidable from the case identifier alone. We show that two of the invariants constrain audit cost and throughput rather than constituting pure operational overhead. We document four architectural patterns from a production Know Your Customer (KYC) deployment for a consumer credit product that reduced multi-day manual processing to same-day automated resolution for approximately four in five cases. Finally, we report three negative results, two identified through internal audit or production observation rather than adversarial testing. We conclude that securing agents in regulated environments depends less on novel attack classes than on rendering auditability, least-privilege authorization, and boundary policy enforcement, with verifiable compliance and bounded audit cost at scale.
\end{abstract}
\bigskip

\section{Introduction}\label{sec:intro}
Large language model (LLM) agents are moving rapidly from experimental assistants into deployed, autonomous components of production software. Unlike a conversational model that answers a question and returns control to a human, an agent perceives an environment, plans, invokes external tools, and takes consequential actions with limited or no human intervention. This transition is now reaching regulated financial systems (consumer credit, payments, fraud detection, and account servicing), where the actor initiating a transaction may now be a non-human agent operating on a person's behalf.

The security research community has responded with a substantial body of work characterizing the agentic attack surface. Layered models decompose the agent stack into foundation, memory, tool, coordination, and governance tiers \cite{chu2026survey}; systematizations catalogue prompt-injection attacks and defenses \cite{wang2026landscape}; and dedicated analyses examine the Model Context Protocol (MCP) tool ecosystem \cite{gaire2025sok}, agent identity and delegated authority \cite{south2025identity}, and memory-state manipulation \cite{xu2026storage}. Collectively this literature establishes that the agentic surface is qualitatively different from that of stateless LLMs: attacks emerge not only at the prompt interface but through persistent state, delegated authority, and long-horizon interaction \cite{chu2026survey}.

Yet this literature shares a structural limitation: it is overwhelmingly laboratory-based. The systematizations are built from literature review and benchmark evaluation, and several state explicitly that they did not deploy or independently replicate in production \cite{gaire2025sok}. In parallel, a separate body of non-peer-reviewed practitioner material (vendor playbooks, compliance-consultant guides, and governance frameworks such as the NIST AI Risk Management Framework and the EU AI Act) addresses how to deploy LLM systems under regulatory regimes such as SOC 2, GDPR, and HIPAA \cite{tabassi2023airmf,iso42001}. No peer-reviewed work addresses the two together, and no prior work has systematized how the academic threat models map onto concrete, legally mandated control obligations of regulated financial deployment, nor reported which mitigations remain effective in production at scale.

When we deployed MCP/RAG/A2A agents to automate Know Your Customer (KYC) onboarding for a consumer credit product, the academic frameworks told us how an agent could be attacked but not what we would be legally obligated to prove after it acted. The taxonomies assumed the operative question was whether an action was authorized; in a regulated consumer-finance setting the binding question was whether we could reconstruct, from a single case identifier, exactly which factors drove an applicant's outcome, and whether every policy the agent applied was the current, approved version. The existing laboratory literature did not anticipate that the most consequential failures in our deployment would be identified not through adversarial testing but through internal audit: a superseded policy in the retrieval store causing systematic over-verification of customers who no longer required it, and tool-call logs that existed but could not be traced to the corresponding decision cases. The gap we encountered was not an unidentified attack class but the absence of a systematic mapping between established threat models and the regulatory control obligations against which our deployment was audited. In this paper, we address this gap.

The main contribution of this paper is not another taxonomy; instead, we offer an experience-grounded systematization in which we (i) map established agentic threat categories onto the specific control obligations imposed by financial regulation; (ii) distill these obligations into five runtime control invariants (C1--C5) and give reference algorithms (Algorithms~\ref{alg:1}--\ref{alg:5}) that realize them at the architectural boundaries a regulated deployment is audited against; (iii) develop a formal model and analysis in which the audit-reconstruction gate is proved sound and complete for the invariants (Theorem~\ref{the:1}), the threat coverage of each invariant is characterized, and the invariants are shown to bound audit cost and throughput; (iv) document how four architectural patterns applied in a production regulated-finance KYC deployment instantiate the control invariants within a formally audited deployment; and (v) report which patterns held under audit and operational load, and which did not, including two identified through internal audit or production observation, not adversarial testing. The invariants, algorithms, and analysis are intended to be reproducible in argument and transferable to other regulated domains.

The remainder of the paper is organized as follows. Section~\ref{sec:background} reviews the agentic-security literature and the regulated deployment guidance and states the gap precisely. Section~\ref{sec:threats} presents a threat model under regulatory constraint and distills it into five runtime control invariants. Section~\ref{sec:algorithms} gives the invariants an operational form as five reference algorithms. Section~\ref{sec:patterns} describes architectural patterns observed in production. Section~\ref{sec:formal} develops a formal model of the invariants, proves the audit-reconstruction gate sound and complete (Theorem~\ref{the:1}), characterizes the threat coverage of each invariant, and derives their cost and throughput bounds. Section~\ref{sec:negative} reports negative results and open problems. Section~\ref{sec:generalizing} discusses generalizations beyond finance, and Section~\ref{sec:conclusions} presents our conclusions.
\section{Background and Related Work}\label{sec:background}
\subsection{Agent Architectures: MCP, RAG, and A2A}\label{sec:arch}

Modern production agents are commonly assembled from three interoperating layers. The Model Context Protocol (MCP) standardizes how an agent discovers and invokes external tools and data sources through a typed and scoped interface \cite{gaire2025sok}. Retrieval-augmented generation (RAG) grounds model output in an external corpus, retrieving relevant documents at inference time to constrain the model to verified information \cite{lewis2020rag}. Agent-to-agent (A2A) orchestration coordinates multiple specialized agents through structured message passing, enabling division of labor across a workflow \cite{google2025a2a}. Though each layer expands capability, they each expand the attack surface in a distinct way simultaneously.

\subsection{Agentic-Security Taxonomies}\label{sec:taxonomies}

Recent systematizations organize the agentic attack surface along structural lines. Layered attack-surface models decompose the stack into foundation, cognitive, memory, tool-execution, multi-agent-coordination, ecosystem, and governance layers, augmented by a temporal axis distinguishing instantaneous, session-persistent, and cross-session threats \cite{chu2026survey}. The prompt-injection literature distinguishes between direct and indirect injection, the latter involving malicious instructions embedded in external content subsequently consumed by the agent \cite{wang2026landscape,greshake2023indirect}. MCP-focused work catalogues tool poisoning, tool-description manipulation, and ``rug-pull'' attacks in which a tool's behavior changes after it is trusted \cite{gaire2025sok}. Memory-centric analyses show that persistent state introduces control-flow hijacking that stateless threat models do not capture \cite{xu2026storage}. Identity-focused work argues that authentication, authorization, and delegated authority constitute a distinct frontier for agentic systems \cite{south2025identity}. Community efforts such as the OWASP Top 10 for LLM Applications consolidate these threat classes into practitioner-facing risk catalogues \cite{owasp2025llm}.

\subsection{Deployment Guidance for Regulated AI}\label{sec:guidance}

A separate body of guidance addresses operating LLM systems under regulatory regimes. Governance frameworks---the NIST AI Risk Management Framework (AI RMF 1.0, 2023), ISO/IEC 42001:2023, and the EU AI Act's high-risk-system obligations define what organizations must document and control \cite{tabassi2023airmf,iso42001}. In US banking, the primary supervisory standard for model governance is the interagency model risk management guidance, most recently revised as SR 26-2 (Federal Reserve, OCC, and FDIC, April 17, 2026), which supersedes SR 11-7 (2011) and SR 21-8 (2021) \cite{frb2026sr262}. Practitioner guides translate these frameworks into architectural prescriptions: VPC isolation, immutable audit logging, data-residency controls, role-based access, and gateway-layer policy enforcement \cite{iso42001}. These prescriptions are organized around compliance requirements rather than around the failure modes of agentic architectures. They enumerate what a compliant system must contain but do not derive those requirements from a model of how an autonomous, multi-agent system can be attacked or can fail. As a result, they remain disconnected from the agentic threat taxonomies of Section~\ref{sec:taxonomies}, which characterize those failure modes but do not map them to regulatory obligations.

\subsection{The Gap}\label{sec:gap}

This disconnect extends to the regulatory frameworks themselves. The academic taxonomies are rigorous but abstract from the binding constraints (legal accountability, audit obligation, data-residency law) that dominate a regulated deployment. The practitioner guidance is concrete but not grounded in a formal threat model. A concrete marker of this gap emerged in April 2026: the revised interagency model risk management guidance (SR 26-2) explicitly excludes generative AI and agentic AI models from its scope, stating that they ``are novel and rapidly evolving'' and deferring their governance to general organizational risk management practices \cite{frb2026sr262}. The primary US bank supervisory framework for model governance does not reach the autonomous agents now entering consumer-facing financial systems.

Two recent papers partially address, but do not fully resolve, this gap. Rashie and Rashi \cite{rashie2026typechecked} encode financial regulatory obligations (SEC Rule 15c3-5, FINRA Rule 3110, OCC Bulletin 2011-12) as formal proof obligations for agentic systems using Lean 4 theorem, but do not perform adversarial threat-model analysis, leaving the question of how attacks exploit the gap between architecture and obligation unaddressed. Maiti \cite{maiti2026caging} performs the bridging operation for healthcare, mapping six threat domains to specific HIPAA Security Rule provisions in a nine-agent production deployment, but does not address the distinct regulatory obligations of financial services. Threat-model frameworks for enterprise agents \cite{narajala2025securing} and multi-agent security taxonomies \cite{schroederdewitt2025open} omit regulatory mapping entirely, and execution-grounded financial safety benchmarks such as FinVault \cite{yang2026finvault} describe compliance constraints generically rather than grounding them in named statutory obligations. No prior work provides a threat-model-grounded security analysis that is simultaneously (a) rooted in adversarially validated attack taxonomies for LLM agents, (b) mapped to the specific legally mandated control obligations of regulated financial deployment, and (c) grounded in observations from a production financial deployment. We integrate these three dimensions into a unified, deployment-grounded framework for analyzing agent security in regulated financial systems.

\begin{figure}[tbp]
\centering
\includegraphics[width=\textwidth]{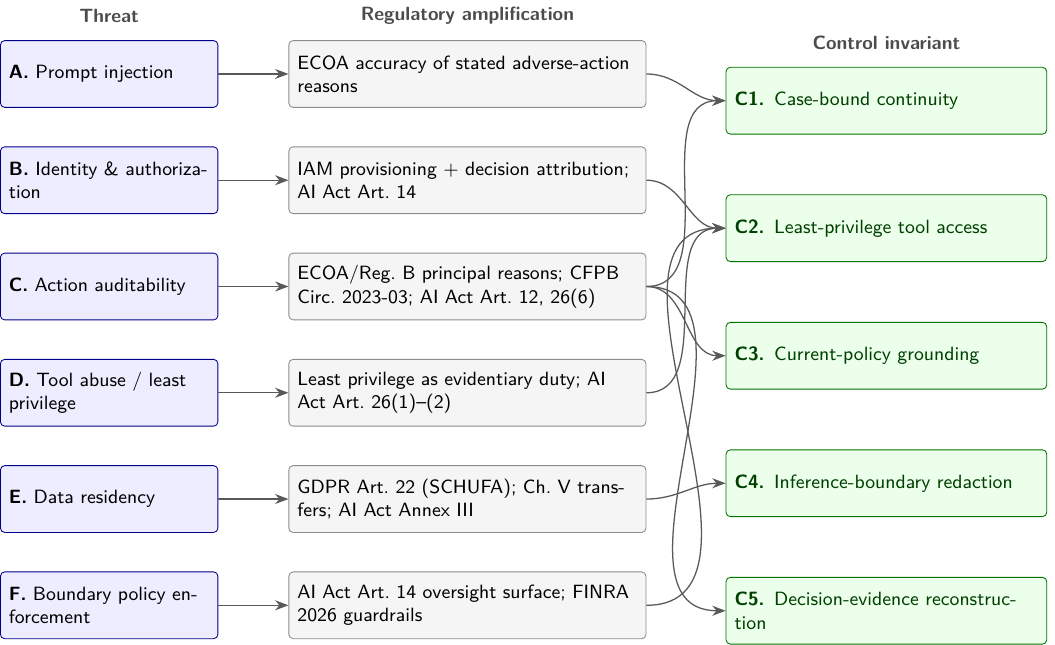}
\caption{Mapping from established agentic threat categories (left; Section~\ref{sec:threats}) through their regulatory amplification (center) to the control invariants of Section~\ref{sec:invariants} that discharge the resulting obligations (right). Sources for the amplification column: ECOA, Regulation~B and CFPB Circular 2023-03 \cite{cfpb2023circular}, the EU AI Act \cite{eu2024aiact}, GDPR Article~22 \cite{cjeu2023schufa} and FINRA's 2026 guidance \cite{finra2026report}. Horizontal links pair each threat with its amplification; invariants C1 (case-bound continuity) and C2 (least-privilege tool access) are each induced by multiple threats.}
\label{fig:1}
\end{figure}
\section{Threat Model Under Regulatory Constraints}\label{sec:threats}
This section examines each established agentic threat category and identifies its regulatory amplification: how legal accountability, audit obligations, and data-protection requirements affect the severity, priority, or required mitigation of the threat relative to an unregulated deployment. The threat categories are drawn from the existing literature \cite{chu2026survey,wang2026landscape,gaire2025sok,south2025identity,xu2026storage}, whereas the analysis of their regulatory amplification, supported by observations from a production deployment, constitutes the contribution of this work. Figure~\ref{fig:1} summarizes the full mapping developed in this section.

\subsection{Prompt Injection Under Accountability}\label{sec:injection}

Prompt injection is the canonical agentic threat: an attacker supplies instructions that the model treats as authoritative, either directly through the user interface or indirectly through external content the agent later consumes \cite{wang2026landscape,greshake2023indirect}. The indirect variant is more dangerous in an automated compliance pipeline, because such pipelines exist precisely to process untrusted external input at scale.

In this deployment, raw documents were not directly exposed to the model. Instead, a structured-extraction stage provided only the required fields after redaction of personally identifiable information (see Section~\ref{sec:residency}). This approach reduces but does not eliminate the indirect prompt-injection attack surface, as a crafted value within an extracted field may still reach the model. This surviving surface is the threat's bounded residual, the portion of an attack surface that a mitigation reduces but does not eliminate, and regulatory amplification is what makes even this bounded residual consequential. An injected value that alters the KYC determination does not merely cause an unauthorized action; it produces an onboarding or adverse-action decision whose actual basis diverges from the factors the institution believes it applied. Under ECOA and Regulation B, 15 U.S.C. \S{}1691(d)(2)--(3) and 12 C.F.R. \S{}1002.9(b)(2) require that adverse-action reasons ``relate to and accurately describe the factors actually considered or scored by a creditor'' \cite{cfpb2023circular}. An injection that silently alters the decision-making process may therefore result in an applicant-specific ECOA adverse-action notice accuracy violation, where the stated reason no longer reflects the actual basis for the decision, irrespective of any downstream harm. In an unregulated chatbot, a successful injection is a security incident that needs to be contained, whereas in a regulated domain, it is a compliance breach that the institution must be able to detect and remediate after the fact, not merely prevent, because the adverse-action notice has already been issued on an inaccurate basis. This shifts the required defensive posture from prevention to prevention plus per-decision attributability, the same requirement that motivates the audit patterns of Section~\ref{sec:patterns}.

\subsection{Identity and Authorization for Non-Human Actors}\label{sec:identity}

In unregulated deployments, agent identity and authorization are primary security concerns as an attacker who can impersonate an agent or escalate its permissions can take unauthorized actions. The research community addresses this through authentication protocols, capability-bounded tool invocation, and delegated-authority frameworks \cite{south2025identity,narajala2025securing}. The regulatory amplification in consumer finance operates on two distinct axes, namely identity provisioning and decision attribution, each of which the academic frameworks do not address.

The first axis is identity provisioning. In a regulated consumer-finance deployment, every actor that touches customer financial data must be provisioned in the institution's identity and access management (IAM) system; this is the assumption embedded in GLBA (Gramm--Leach--Bliley Act) Safeguards Rule for vendor management controls and in the access-review evidence that SOC 2 audits demand. Autonomous agents do not readily conform to this model. They require machine credentials (service accounts or OAuth client identifiers), but current production IAM infrastructures have no workflows for scoping those credentials to the bounded capability set of a particular agent's intended function. The operational default became production-level service accounts whose permissions far exceeded what any single agent task required. The regulatory consequence was direct and audit-visible: access reviews found service-account grants for which no role justification existed, because the role-based model assumed a human employee with defined job duties, not an autonomous agent with a bounded and time-limited task. Over-privilege in that context is not merely a security risk; it is a control failure that produces adverse audit findings independent of whether any attack has occurred.

The second axis is decision attribution. Academic literature treats agent identity as an authorization boundary problem; in a production regulated-finance environment it becomes an attribution problem with statutory consequences. ECOA adverse-action requirements (15 U.S.C. \S{}1691(d)(2)--(3); 12 C.F.R. \S{}1002.9(b)(2)) require a traceable decision-maker whose factors can be specifically disclosed to the consumer \cite{cfpb2023circular}. When an orchestrating agent delegates retrieval to a sub-agent whose output influences a downstream credit outcome, the decision-maker in the audit trail becomes ambiguous across agent boundaries, and that ambiguity itself becomes a regulatory exposure. The consumer is owed a specific principal reason grounded in the factors considered; the workflow should return a score and without attributing that score to an identifiable, bounded decision process cannot satisfy this statutory obligation regardless of the score's accuracy.

On the EU side, AI Act Article 14(1) translates these concerns into a design obligation: high-risk AI systems must be built so that natural persons can effectively oversee them while in use \cite{eu2024aiact}. Article 14(3) scales the obligation with the system's level of autonomy: a more autonomous agent triggers proportionately stronger oversight requirements, implementable by the provider before market placement or by the deployer after receipt. Article 14(4) further requires that designated overseers hold actual authority to halt the system in any situation. An agent operating under a non-human identity with no human principal empowered to stop it fails Article 14 independently of any adversarial action, making the identity and authorization architecture a compliance artifact as well as security control.

\subsection{Action Auditability as a Legal Obligation}\label{sec:auditability}

In unregulated deployments, audit logging is an operational concern: logs support debugging, anomaly detection, and incident response. The security literature treats auditability as a defense-in-depth property, useful for forensics after a compromise but not itself a threat category \cite{chu2026survey,narajala2025securing}. In regulated consumer finance, this framing inverts. Action auditability is not a monitoring best practice; it is a statutory obligation whose violation is independently actionable regardless of whether any security event has occurred. 

The first axis of amplification is the legal-evidence reframing of what a log must be. ECOA (15 U.S.C. \S{}1691(d)(2)--(3)) and Regulation B (12 C.F.R. \S{}1002.9(a)(2)(i), \S{}1002.9(b)(2)) require that any adverse action notice specifically state the principal reason or reasons for the action and accurately describe the factors considered or scored by the creditor. CFPB Circular 2023-03 (September 19, 2023) stated that creditors using AI or complex credit models may not substitute the sample-form checklist of reasons unless those reasons specifically and accurately reflect the actual factors relied upon \cite{cfpb2023circular}. The Bureau withdrew the Circular, together with other sub-regulatory guidance, as of May 12, 2025 \cite{cfpb2025withdrawal}, but the requirement it interpreted is in Regulation B itself (12 C.F.R. \S{}1002.9(b)(2)) and was left unchanged by the 2026 amendments to Regulation B \cite{cfpb2026regb}. FCRA adds parallel disclosure obligations when a credit score from a consumer reporting agency contributes to the outcome. The aggregate effect is that the audit trail for any agent action touching a credit outcome must be capable of reconstructing the principal reasons in a consumer-disclosable form, a requirement qualitatively different from operational telemetry. Most agent frameworks produce logs adequate for debugging; in a production regulated-finance deployment, none natively produced a per-decision evidence record that could satisfy an adverse-action disclosure obligation without additional reconstruction infrastructure built on top.

The second axis is the structural challenge that chained-agent architectures posed to that reconstruction. In a single-model system, the decision is a single inference call whose inputs (prompt, context, parameters) can be captured as one unit. In an MCP/RAG/A2A pipeline, the decision is distributed: a retrieval agent fetches documents, a reasoning agent synthesizes them, and an orchestrator aggregates the result into an outcome. The audit question is not merely ``what did the system output?'' but ``which retrieved documents contributed which factors to which outcome, and in what proportion?'' Answering that question in a form that satisfies the statutory standards required instrumenting the inter-agent message boundaries, not the model calls, as the primary unit of audit evidence. This is a non-obvious architectural implication that does not appear in any of the surveyed laboratory threat models, which treat the agent as a unit rather than a pipeline of attributable sub-decisions.

EU AI Act Article 12(1) codifies this obligation on the provider side: high-risk AI systems ``shall technically allow for the automatic recording of events (logs) over the lifetime of the system'' \cite{eu2024aiact}. Article 12(2) further specifies the purposes those logs must serve: traceability appropriate to the system's intended purpose; identification of risk situations under Article 79(1); post-market monitoring under Article 72; and deployer operational monitoring under Article 26(5). Article 26(6) separately requires deployers to retain automatically generated logs for at least six months. The combined effect of Articles 12 and 26 is that auditability is a design requirement binding on the provider and a retention requirement binding on the deployer, neither of which is satisfiable by general-purpose observability tooling that was not designed with statutory evidence reconstruction in mind.

\subsection{Tool Abuse and Least Privilege Under Audit}\label{sec:toolabuse}

The tool-execution layer is where agentic systems take consequential action, and the security literature accordingly treats tool abuse (privilege escalation, unauthorized invocation, capability misuse) as a primary threat to be contained through least-privilege bounding \cite{chu2026survey,gaire2025sok}. In an unregulated deployment, least-privilege is a hardening recommendation: it reduces the scope of impact of a compromise. Regulation converts it into an evidentiary obligation.

In the KYC pipeline, the orchestrator invoked MCP tools for bureau retrieval, document verification, policy lookup, and ACH checks. As described in Section~\ref{sec:identity}, these tools were initially provisioned with broad service-account permissions because the IAM model had no construct for binding a credential to a single agent's task scope. The audit-relevant finding was not that any tool had been abused, but that the agent could invoke capabilities beyond what an individual KYC case required, and that this latent over-privilege had no role justification in the access review. The distinction is the amplification: in a regulated setting it is not sufficient to bound and document what the agent did and log it; one must be able to demonstrate to an auditor that the agent could not have exceeded its mandate. EU AI Act Article 26(1) codifies the deployer's side of this obligation, requiring ``appropriate technical and organizational measures to ensure they use such systems in accordance with the instructions for use,'' and Article 26(2) requires that human oversight be assigned to persons with ``the necessary competence, training and authority'' \cite{eu2024aiact}. An agent holding access to tools beyond those its instructions for use contemplate is therefore in breach of the deployer obligation before any attack occurs: least-privilege becomes a precondition of compliant operation, not a plausible defense-in-depth feature.

\subsection{Data Residency and Sovereignty Constraints}\label{sec:residency}

Data residency is largely absent from the agentic-security taxonomies, which treat the model as a fixed computational resource rather than as a network endpoint whose physical location carries legal consequence. In a regulated deployment, where the data is processed is in itself a controlled property. Even within a single jurisdiction, the agent architecture introduced data-egress points that the rules engine it replaced did not have: customer financial data now transited through a RAG retrieval store and a model inference endpoint, each a point at which regulated data could cross the boundary of the environment authorized to handle it. Establishing where customer data was processed at each hop and being able to prove that it stayed within the permitted boundary became a precondition for deployment rather than an operational afterthought.

The amplification is sharpest at the inference boundary, because this model endpoint is the hardest hop to constrain: it is frequently managed by a third-party service, often in a different jurisdiction, and it consumes whatever context the agent sends it. The pattern we adopted to bound this exposure was to never send raw customer documents to the model at all. Documents were processed through a structured-extraction stage that pulled only the specific fields the KYC determination required, and those fields passed through a redaction proxy that removed or tokenized personally identifiable information before any data reached the inference endpoint. The model reasoned over redacted, structured inputs; the mapping from tokens back to identities never left the controlled environment. This kept the inference hop free of resident PII, narrowing the residency obligation to extraction and proxy components, which were sited deliberately within the permitted boundary.

The regulatory grounding makes the consequence of non-compliance concrete. The CJEU's SCHUFA judgment (Case C-634/21, ECLI:EU:C:2023:957) established that GDPR Article 22(1) applies to AI-based credit scoring when a third party ``draws strongly'' on the resulting value to establish or terminate a contractual relationship \cite{cjeu2023schufa}, bringing the entire processing pipeline including any cross-border transfer to non-EU inference infrastructure within the GDPR framework. A pipeline that routed an EU-resident's unredacted financial data to a non-EEA model endpoint without a Chapter V transfer safeguard would violate the transfer rules, and if the output drove a credit denial, it meant that Article 22 is violated as well. Under EU AI Act Annex III, point 5(b), creditworthiness evaluation systems are classified as high-risk (with a carve out enabled only for pure fraud detection), so these residency obligations are further amplified by the full suite of Annex III provider and deployer duties \cite{eu2024aiact}, which apply from 2 December 2027 after the 2026 AI Omnibus deferred their original August 2026 date \cite{eu2026omnibus}. The redaction-proxy pattern is, in effect, a residency control: by ensuring that what crosses the hardest-to-constrain boundary is no longer personal data, it reduces the surface over which these obligations bind.

\subsection{Policy Enforcement at the Request Boundary}\label{sec:boundary}

Conventional cloud security relies on a principle that agentic architectures do not preserve: policy is enforced at the boundary. In a network-security posture, every consequential request traverses a control point (a gateway or proxy) where it is inspected, authorized, and logged before it reaches the resource it targets, and the workload cannot bypass that point. Advancements in the cloud security discipline in the last decade are predominantly aimed at making the boundary comprehensive and unforgeable.

Agent architectures dissolve this boundary. The consequential action is no longer an outbound request that crosses a gateway; it is a tool call emitted mid-reasoning, internal to the agent's execution, with no inherent enforcement point between the model's decision and the tool's effect. Reconstructing a boundary therefore required choosing one deliberately. We treated the MCP tool-invocation layer as the new policy enforcement point: the single place where a request could be inspected, authorized against the agent's mandate, and logged with the originating case identifier (Section~\ref{sec:caseid}). The regulatory amplification is that this one boundary must now satisfy three regimes simultaneously. EU AI Act Article 14(1)--(3) requires that high-risk systems be designed so humans can effectively oversee them, with oversight proportionate to autonomy \cite{eu2024aiact}, an obligation that presupposes an interceptable enforcement point. FINRA's 2026 Annual Regulatory Oversight Report, the first to treat AI agents as a distinct supervisory category, identifies ``implementing guardrails to constrain or restrict AI agent behaviors, actions, or decisions'' as a required supervisory consideration and holds that ``firms remain responsible for recommendations and decisions influenced by GenAI'' regardless of autonomy \cite{finra2026report}. And the ECOA audit obligations of Section~\ref{sec:auditability} require that whatever crosses the boundary be logged in a form that can later reconstruct the decision. The tool-invocation boundary is thus simultaneously a security control, an EU AI Act oversight surface, and a FINRA-mandated guardrail: three regimes converging on a single architectural seam that the agent frameworks do not expose by default.

\subsection{Control Invariants Induced by the Threat Mapping}\label{sec:invariants}

The regulatory amplifications of Sections~\ref{sec:injection} through~\ref{sec:boundary} converge on a small set of properties that must hold for every case, independent of which threat is active. We state them as five runtime \textbf{control invariants}. They bridge the mapping of Figure~\ref{fig:1} to the reference algorithms of Section~\ref{sec:algorithms} and the formal analysis of Section~\ref{sec:formal}, and each is enforced at an identifiable architectural component. Figure~\ref{fig:3}, presented with the deployment patterns of Section~\ref{sec:patterns}, locates every invariant on the component that enforces it.

\textbf{C1 --- Case-Bound Continuity.} Every consequential action --- tool call, policy lookup, inference, and decision --- carries the same opaque case identifier, the precondition for per-decision attributability (Section~\ref{sec:injection}, Section~\ref{sec:auditability}).

\textbf{C2 --- Least-Privilege Tool Access.} An agent may invoke only the tools and capabilities explicitly allowed for its assigned workflow and role; regulation makes this an evidentiary obligation, not only hardening (Section~\ref{sec:identity}, Section~\ref{sec:toolabuse}).

\textbf{C3 --- Current-Policy Grounding.} Every policy-dependent action uses the currently approved and effective policy version; a superseded provision is a control failure even when it over-verifies (Section~\ref{sec:auditability}; the failure of Section~\ref{sec:stalepolicy} is a C3 violation).

\textbf{C4 --- Inference-Boundary Redaction.} No raw document or resident PII crosses the inference boundary; only structured, redacted fields reach the model, bounding both injection residual (Section~\ref{sec:injection}) and residency exposure (Section~\ref{sec:residency}).

\textbf{C5 --- Decision-Evidence Reconstruction.} From the case identifier alone, the entire decision basis --- tools, policy, redaction, and factors --- must be reconstructable in a consumer-disclosable form (Section~\ref{sec:auditability}).

Two of these --- C1 (auditability) and C2 (least-privilege) --- are each induced by multiple threats (Figure~\ref{fig:1}) and, as Section~\ref{sec:cost} shows, also govern the performance envelope, so they are performance controls as much as compliance ones.
\section{Reference Algorithms for the Control Invariants}\label{sec:algorithms}

We give the invariants of Section~\ref{sec:invariants} an operational form as five reference algorithms, framework-agnostic and annotated with the invariant each guarded step discharges. Algorithm~\ref{alg:1} is the top-level workflow; Algorithms~\ref{alg:2}--\ref{alg:5} are the invariant-bearing subroutines it calls.


\begin{algorithm}[!htbp]
\caption{Execute\_Regulated\_Workflow}\label{alg:1}
\begin{algorithmic}[1]
\Require \emph{request}
\Ensure \emph{decision} or \emph{human\_review\_reference}
\State $\mathit{case\_id} \gets \Call{Init\_Case\_And\_Context}{\mathit{request}}$ \Comment{mint opaque Case ID (C1)}
\Try
  \State $\mathit{redacted}, \mathit{red\_ev} \gets \Call{Redact\_Before\_Inference}{\mathit{case\_id}, \mathit{request.input}}$ \Comment{C4, Alg.~\ref{alg:4}}
  \State $\mathit{policy\_ev} \gets \Call{Retrieve\_Current\_Policy}{\mathit{case\_id}, \mathit{request.workflow}, \mathit{request.jurisdiction}}$ \Comment{C3, Alg.~\ref{alg:3}}
  \State $\mathit{plan} \gets \Call{Orchestrator.Plan}{\mathit{request.workflow}, \mathit{policy\_ev}, \mathit{redacted}}$
  \Statex \Comment{model calls receive only the redacted payload and log an INFERENCE event (C4)}
  \State $\mathit{results} \gets \emptyset$
  \For{\textbf{each} $\mathit{step} \in \mathit{plan.steps}$} \Comment{steps drawn from the C2-allowed capability set}
    \State $\mathit{result} \gets \Call{Invoke\_Controlled\_Tool}{\mathit{case\_id}, \mathit{step}, \mathit{redacted}, \mathit{policy\_ev}}$ \Comment{C2, Alg.~\ref{alg:2}}
    \State $\mathit{results}.\mathrm{append}(\mathit{result})$
  \EndFor
  \State $\mathit{decision}, \mathit{factors} \gets \Call{Decision\_Engine.Decide}{\mathit{case\_id}, \mathit{results}, \mathit{policy\_ev}}$
  \State \Call{Audit\_Log}{$\mathit{case\_id}$, DECISION\_RECORDED, $\mathit{decision}$, $\mathit{factors}$}
  \Statex \Comment{C1, C5: decision event enters the ledger before reconstruction}
\Catch{ControlException} \Comment{any raise from Algs.~\ref{alg:2}--\ref{alg:4}}
  \State \Return \Call{Route\_To\_Human\_Review}{$\mathit{case\_id}$, CONTROL\_EXCEPTION}
\EndTry
\State $\mathit{status} \gets \Call{Reconstruct\_Decision\_Evidence}{\mathit{case\_id}}$ \Comment{C5, Alg.~\ref{alg:5}}
\If{$\mathit{status} \neq \mathrm{COMPLETE}$}
  \State \Return \Call{Route\_To\_Human\_Review}{$\mathit{case\_id}$, INCOMPLETE\_EVIDENCE}
\EndIf
\State \Call{Persist\_Decision\_Outcome}{$\mathit{case\_id}, \mathit{decision}, \mathit{factors}$}
\State \Return $\mathit{decision}$
\end{algorithmic}
\end{algorithm}

\begin{algorithm}[!htbp]
\caption{Invoke\_Controlled\_Tool}\label{alg:2}
\begin{algorithmic}[1]
\Require \emph{case\_id}; \emph{step}; \emph{payload}; policy evidence \emph{policy\_ev}
\Ensure \emph{tool\_response}, or a raised control exception
\State $\mathit{env} \gets \Call{Build\_Tool\_Envelope}{\mathit{case\_id}, \mathit{step}, \textproc{Hash}(\mathit{payload}), \mathit{policy\_ev.ref}}$
\Statex \textbf{Authorization gate}
\If{\textbf{not} \Call{Authorization\_Engine.Authorize}{$\mathit{step.agent}, \mathit{step.tool}, \mathit{step.capability}, \mathit{case\_id}$}} \Comment{C2}
  \State \Call{Audit\_Log}{$\mathit{case\_id}$, TOOL\_AUTHZ\_DENIED, $\mathit{env}$}
  \State \textbf{raise} AuthorizationError
\EndIf
\Statex \textbf{Inference-boundary gate}
\If{\textbf{not} \Call{Boundary\_Enforcer.Allow}{$\mathit{env.boundary\_class}$}} \Comment{C4}
  \State \Call{Audit\_Log}{$\mathit{case\_id}$, BOUNDARY\_VIOLATION, $\mathit{env}$}
  \State \textbf{raise} BoundaryViolationError
\EndIf
\Statex \textbf{Policy-grounding gate}
\If{\textbf{not} \Call{Policy\_Grounding\_Valid}{$\mathit{policy\_ev}$}} \Comment{C3}
  \State \Call{Audit\_Log}{$\mathit{case\_id}$, POLICY\_GROUNDING\_FAILED, $\mathit{env}$}
  \State \textbf{raise} PolicyGroundingError
\EndIf
\Statex \textbf{Invoke and record}
\State \Call{Audit\_Log}{$\mathit{case\_id}$, TOOL\_INVOKED, $\mathit{env}$} \Comment{C1: invocation recorded before execution}
\State $\mathit{response} \gets \Call{Tool\_Adapter.Invoke}{\mathit{step.tool}, \mathit{env}}$
\State \Call{Audit\_Log}{$\mathit{case\_id}$, TOOL\_COMPLETED, \textproc{Hash}($\mathit{payload}$), \textproc{Hash}($\mathit{response}$), $\mathit{step.tool}$, $\mathit{policy\_ev.ref}$} \Comment{C1, C5}
\State \Return $\mathit{response}$
\end{algorithmic}
\end{algorithm}

\begin{algorithm}[!htbp]
\caption{Retrieve\_Current\_Policy}\label{alg:3}
\begin{algorithmic}[1]
\Require \emph{case\_id}; \emph{workflow}; \emph{jurisdiction}
\Ensure \emph{policy\_evidence}
\State $\mathit{cand} \gets \Call{Policy\_Store.Query}{\mathit{workflow}, \mathit{jurisdiction}}$
\State $\mathit{approved} \gets \Call{Filter}{\mathit{cand}, \mathit{state} = \mathrm{APPROVED}}$
\State $\mathit{current} \gets \Call{Filter}{\mathit{approved}, \mathit{effective\_date} \le \mathrm{now} \le \mathit{expiry\_date}}$
\If{$\mathit{current} = \emptyset$}
  \State \Call{Audit\_Log}{$\mathit{case\_id}$, NO\_CURRENT\_POLICY}
  \State \textbf{raise} NoCurrentPolicyError
\EndIf
\State $\mathit{selected} \gets \Call{Select\_Highest\_Version}{\mathit{current}}$
\If{\textbf{not} \Call{Verify\_Digest}{$\mathit{selected.digest}$}}
  \State \Call{Audit\_Log}{$\mathit{case\_id}$, POLICY\_INTEGRITY\_FAILED}
  \State \textbf{raise} PolicyIntegrityError
\EndIf
\State $\mathit{evidence} \gets \Call{Create\_Policy\_Evidence}{\mathit{case\_id}, \mathit{selected}}$
\State \Call{Audit\_Log}{$\mathit{case\_id}$, POLICY\_RETRIEVED, $\mathit{evidence.ref}$}
\State \Return $\mathit{evidence}$
\end{algorithmic}
\end{algorithm}

\begin{algorithm}[!htbp]
\caption{Redact\_Before\_Inference}\label{alg:4}
\begin{algorithmic}[1]
\Require \emph{case\_id}; \emph{raw\_input}
\Ensure \emph{sanitized\_payload}; \emph{redaction\_evidence}
\State $\mathit{fields} \gets \Call{Structured\_Extraction}{\mathit{raw\_input}}$
\State $\mathit{pii} \gets \Call{Detect\_PII}{\mathit{fields}}$
\If{$\mathit{pii.contains\_pii}$}
  \State $\mathit{sanitized} \gets \Call{Redact\_Or\_Tokenize}{\mathit{fields}, \mathit{pii}}$
\Else
  \State $\mathit{sanitized} \gets \mathit{fields}$
\EndIf
\State $\mathit{sanitized} \gets \Call{Remove\_Free\_Text}{\mathit{sanitized}}$
\State $\mathit{bclass} \gets \Call{Classify\_Boundary}{\mathit{sanitized}}$
\If{\textbf{not} \Call{Boundary\_Enforcer.Allow}{$\mathit{bclass}$}}
  \State $\mathit{evidence} \gets \Call{Create\_Redaction\_Evidence}{\mathit{case\_id}, \mathit{bclass}, \mathrm{DENY}}$
  \State \Call{Audit\_Log}{$\mathit{case\_id}$, REDACTION\_BOUNDARY\_FAILED, $\mathit{evidence.ref}$}
  \State \textbf{raise} InferenceBoundaryError
\EndIf
\State $\mathit{evidence} \gets \Call{Create\_Redaction\_Evidence}{\mathit{case\_id}, \mathit{bclass}, \mathrm{ALLOW}}$
\State \Call{Audit\_Log}{$\mathit{case\_id}$, REDACTION\_COMPLETED, $\mathit{evidence.ref}$}
\State \Return $\mathit{sanitized}, \mathit{evidence}$
\end{algorithmic}
\end{algorithm}

\begin{algorithm}[!htbp]
\caption{Reconstruct\_Decision\_Evidence}\label{alg:5}
\begin{algorithmic}[1]
\Require \emph{case\_id}
\Ensure $\mathit{status} \in \{\mathrm{COMPLETE}, \mathrm{INCOMPLETE}, \mathrm{FAILED}\}$
\State $\mathit{events} \gets \Call{Audit\_Ledger.Fetch\_By\_Case}{\mathit{case\_id}}$ \Comment{$O(E_{\mathit{case}})$ via C1 index}
\State \textbf{if} $\mathit{events} = \emptyset$ \textbf{then return} FAILED
\State \textbf{if} \Call{Has\_Sequence\_Gaps}{$\mathit{events}$} \textbf{or} \Call{Exists\_Malformed\_Event}{$\mathit{events}$} \textbf{then return} FAILED
\Statex \Comment{C1: missing events surface as sequence gaps; mis-tagged events are excluded by A1}
\State $\mathit{chain} \gets \Call{Build\_Evidence\_Chain}{\mathit{events}}$
\State \textbf{if not} $\mathit{chain.has\_policy}$ \textbf{then return} INCOMPLETE \Comment{C3}
\State \textbf{if} \Call{Inference\_Used}{$\mathit{events}$} \textbf{and} (\textbf{not} $\mathit{chain.has\_redaction}$ \textbf{or}
\Statex \hspace{\algorithmicindent}\Call{Some\_Inference\_Payload\_Unredacted}{$\mathit{events}$}) \textbf{then return} FAILED
\Statex \Comment{C4: fails if any inference event lacks a redaction record or carries non-redacted PII}
\State \textbf{if not} $\mathit{chain.has\_decision}$ \textbf{or} $\mathit{chain.tool\_chain} = \emptyset$ \textbf{then return} INCOMPLETE
\State \textbf{if not} \Call{Is\_Chain\_Traversable}{$\mathit{chain}$} \textbf{then return} INCOMPLETE
\State \textbf{if not} \Call{All\_Control\_Invariants\_Satisfied}{$\mathit{chain}$} \textbf{then return} INCOMPLETE \Comment{C1--C5}
\State \Return COMPLETE
\end{algorithmic}
\end{algorithm}

\textbf{Cross-references.} Algorithm~\ref{alg:1} calls Algorithms~\ref{alg:2}--\ref{alg:5}; the marginal $\triangleright$ comments name the invariant each guarded step discharges, matching the badges in Figure~\ref{fig:3} and the predicates of Section~\ref{sec:model}. Status semantics: FAILED indicates that the integrity of the chain is broken or an invariant is violated; INCOMPLETE indicates that required evidence is missing without a proven violation. Every event carries a per-case monotonic sequence number assigned at append time, so missing events surface as sequence gaps (Algorithm~\ref{alg:5}, line~3); an event appended under the wrong case, or not appended at all, is a gate bypass that Assumption A1 excludes rather than one line~3 detects (Section~\ref{sec:correctness}). Model inference occurs only inside Orchestrator.Plan and Decision\_Engine.Decide; both receive only the redacted payload and log an INFERENCE event carrying the redaction-evidence reference, populating the inference events checked at Algorithm~\ref{alg:5}, line~6.
\section{Architectural Patterns Observed in Production}\label{sec:patterns}
\begin{figure}[tbp]
\centering
\includegraphics[width=\textwidth]{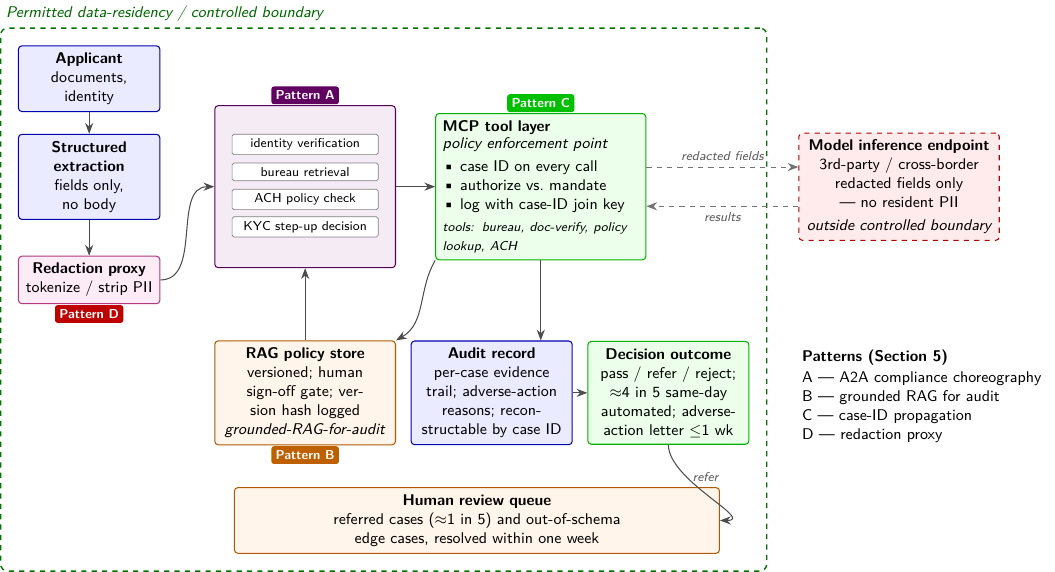}
\caption{Deployment architecture of the KYC onboarding pipeline. Customer documents are reduced to extracted fields and stripped of PII by a redaction proxy (Pattern D) before any data crosses the controlled boundary to the third-party inference endpoint. An orchestrating agent (Pattern A) coordinates four sub-agents through an MCP tool layer that serves as the policy-enforcement point, propagating a case identifier on every call and logging it as the audit join key (Pattern C); the same layer is the enforcement seam for boundary policy (Threat F), not a separate pattern. Policy retrieval is grounded in a versioned, human-approved RAG store (Pattern B). Referred cases and out-of-schema edge cases route to human review.}\label{fig:2}
\end{figure}

In this section, we describe four architectural patterns applied in a production deployment of MCP/RAG/A2A agents supporting Know Your Customer (KYC) compliance for a credit product in US regulated consumer finance. Each pattern is tied to a threat-model category from Section~\ref{sec:threats} and reported at the architectural level. The baseline against which outcomes are measured was a mixed manual-and-rules-based process handling on the order of 100 cases per day at a median resolution time of roughly two to three days per case. Figure~\ref{fig:2} shows how these four patterns compose into the end-to-end KYC pipeline, from customer-document intake through a pass/refer/reject outcome; each component is annotated with the pattern that governs it. We label the four patterns A--D in the order presented below, and Figure~\ref{fig:2} and Table~\ref{tab:1} use these labels; the MCP tool layer through which Patterns C and D operate is shown as a distinct boundary-enforcement seam rather than as a fifth pattern. Table~\ref{tab:1} summarizes each pattern's threat, obligation, mechanism, and observed outcome.

\begin{figure}[tbp]
\centering
\includegraphics[width=\textwidth]{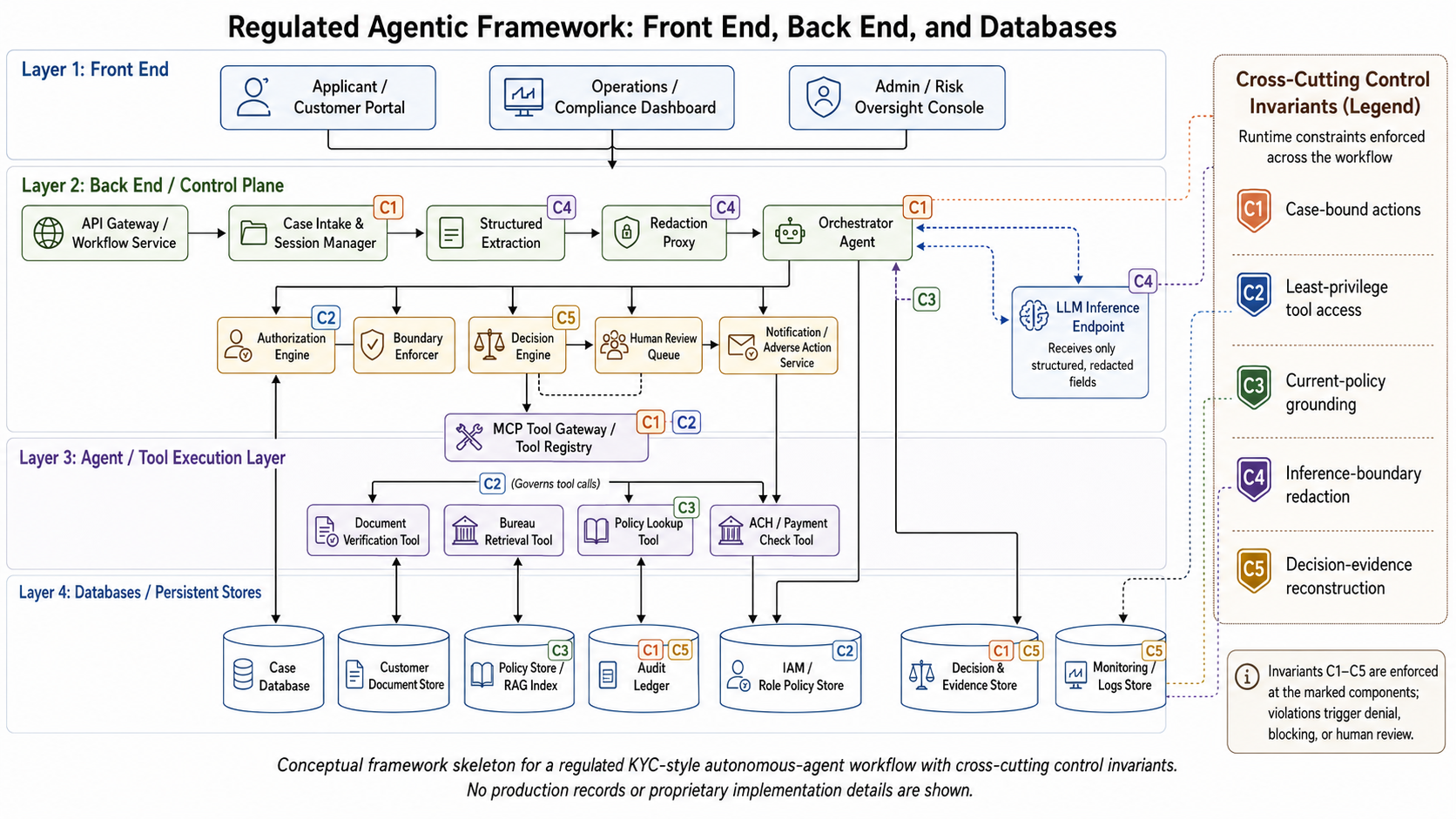}
\caption{Regulated agentic framework across four layers (front end, back-end control plane, agent/tool execution, and persistent stores). Each component is annotated with the control invariant (C1--C5) enforced there; the legend restates the invariants. No production records or proprietary implementation details are shown.}\label{fig:3}
\end{figure}

Whereas Figure~\ref{fig:2} traces a single case through the pipeline, \textbf{Figure~\ref{fig:3}} presents the same deployment as a layered reference architecture and is the figure to which the control invariants are anchored. It separates the system into four layers --- the front end through which applicants and reviewers interact; the back-end control plane that holds the orchestrator, policy service, and authorization engine; the agent and tool execution layer in which MCP tool calls are issued; and the persistent stores that retain case context, policy versions, and audit evidence. Each component carries a badge naming the invariant enforced at that point, so the figure reads as a map from C1--C5 to the components responsible for them: case identity is minted in the control plane (C1), tool authorization is evaluated at the MCP boundary (C2), policy grounding is resolved against the versioned store (C3), redaction is applied before the inference boundary is crossed (C4), and the evidence required for reconstruction is written to the audit and decision stores (C5). The layered view makes explicit that no single component enforces compliance on its own; the guarantee proved in Theorem~\ref{the:1} depends on the tool-invocation boundary being the only path by which the execution layer reaches an external service.

\begin{table}[tbp]
\centering
\caption{Architectural patterns observed in production, each mapped to the threat it addresses (Section~\ref{sec:threats}), the control obligation it satisfies (Figure~\ref{fig:1}), its mechanism, and its observed outcome. Boundary policy enforcement (Threat F) is realized by the MCP tool layer that Patterns C and D operate through and is therefore not listed as a separate row.}\label{tab:1}
\begin{tabularx}{\textwidth}{@{}>{\raggedright\arraybackslash}p{0.15\textwidth}>{\raggedright\arraybackslash}p{0.14\textwidth}>{\raggedright\arraybackslash}p{0.11\textwidth}YY@{}}
\toprule
\textbf{Pattern} & \textbf{Threat (Section~\ref{sec:threats})} & \textbf{Obligation} & \textbf{Mechanism} & \textbf{Observed Outcome} \\
\midrule
A --- A2A Compliance Choreography & C: Action auditability & C1 & Orchestrator drives a deterministic sub-agent sequence; each emits structured output assembled into a pass/refer/reject record with reasons. & $\approx$4/5 same-day straight-through; $\approx$1/5 to human review $\le$1 wk. Structured trail satisfied adverse-action documentation in internal review. \\
\addlinespace
B --- Grounded RAG for Audit & C: Auditability; D: Tool abuse & C3 & Policy docs ingested via compliance feed behind a human sign-off gate; retrieval-time version check records doc ID, version hash, timestamp. & Every step-up decision grounded in a versioned, approved policy. Sign-off gate adds a propagation lag (root cause of failure, Section~\ref{sec:stalepolicy}). \\
\addlinespace
C --- Case-ID Propagation & C: Auditability; B: Identity/authz & C1, C2 & Every MCP tool call carries the originating Case ID as a mandatory header; responses logged with Case ID as the audit join key. & Full tool-call chain reconstructable from Case ID alone. Required retrofitting each tool's interface contract (cost noted, Section~\ref{sec:mcpaudit}). \\
\addlinespace
D --- Redaction Proxy & A: Prompt injection; E: Data residency & C4 & Raw docs never sent to the model; structured extraction + PII tokenization before the inference hop; token$\rightarrow$identity map stays in-boundary. & Injection surface narrowed to extracted fields; residency obligation removed from 3rd-party endpoint. Cost: schema-bound context, edge cases $\rightarrow$ review. \\
\bottomrule
\end{tabularx}
\end{table}

\subsection{A2A Compliance Choreography}\label{sec:a2a}

\textit{Problem} (Section~\ref{sec:auditability}: Action Auditability): The pre-agent KYC process combined automated rules with a human review queue. Because human reviewers operated across multiple systems and produced unstructured notes, the per-case audit trail was inconsistent: some cases could reconstruct the factors driving a refer or reject recommendation; others could not. This created a recurring adverse-action notice compliance risk: ECOA requires that stated reasons accurately describe the factors considered, and an unstructured human review record does not satisfy that standard reliably at volume.

\textit{Mechanism}: An orchestrating agent was introduced to coordinate a deterministic sequence of sub-agent actions: identity document verification, bureau data retrieval, ACH policy applicability check, and KYC step-up determination. Each sub-agent produced a structured output that fed the next stage, with the orchestrator responsible for assembling the final pass/refer/reject recommendation and its supporting reasons. Human reviewers remained in the loop for referred cases; the agent handled straight-through cases end-to-end. Physical adverse-action letters were dispatched within one week for all non-approved outcomes.

\textit{Observed outcome}: Roughly four in five cases resolved same-day through the automated path; the remaining one in five entered a human review queue and completed within one week. Against the pre-agent baseline of roughly two to three days median for all cases, same-day resolution for the majority of volume represents a quantitative shift in the timeliness of credit decisions. The automated path scaled horizontally without adding review staff---its throughput ceiling was set by KYC API rate limits and cost, not headcount---while the human-review queue grew in proportion to the roughly one-in-five referral rate rather than with total volume. The audit trail produced by the orchestrator's structured output consistently satisfied adverse-action documentation requirements in subsequent internal reviews.

\subsection{Grounded RAG for Audit}\label{sec:rag}

\textit{Problem} (Section~\ref{sec:auditability}: Action Auditability; Section~\ref{sec:toolabuse}: Tool Abuse): The KYC and ACH policy documents governing which verification steps applied to a given customer were maintained by a compliance team and updated periodically. Without a controlled retrieval mechanism, the orchestrating agent had no reliable way to determine whether a policy provision was current, superseded, or inapplicable to the specific product. Applying a superseded policy provision (even in the direction of over-verification) constitutes a control failure under SOC 2 access and change-management requirements, because the agent is acting on information that does not accurately reflect the institution's current compliance posture.

\textit{Mechanism}: Policy documents were ingested into a retrieval store via an automated compliance team feed, but with a human sign-off gate introduced before any document version entered the production index. At retrieval time, a policy-version check confirmed that the document being used matched the current approved version rather than a prior ingestion. Each tool call that retrieved a policy document recorded the document identifier, version hash, and retrieval timestamp as part of the case audit record, making the policy basis for each step-up determination explicitly traceable.

\textit{Observed outcome}: The pattern held up under normal operating conditions and produced an audit trail in which every KYC step-up decision was grounded in a specific, versioned, human-approved policy document. The human sign-off gate introduced a processing lag when policies changed---compliance updates could not take effect in the agent until a reviewer approved the new document version---but this lag was accepted as a deliberate control rather than a defect. The pattern did not prevent the failure described in Section~\ref{sec:stalepolicy}; it was the remediation that followed that failure.

\subsection{Case-ID Propagation Through MCP Tool Calls}\label{sec:caseid}

\textit{Problem} (Section~\ref{sec:auditability}: Action Auditability; Section~\ref{sec:identity}: Identity and Authorization): As described in Section~\ref{sec:auditability}, MCP tool calls in the initial deployment produced logs that were adequate for operational debugging but were not correlated to the case identifier that anchored the KYC decision in the institution's system of record. An adverse-action audit that started from a case ID could not reliably traverse back through the MCP tool call logs to reconstruct which document retrievals and policy checks had driven the outcome for that specific applicant. The logs existed; the linkage did not.

\textit{Mechanism}: A case-ID propagation layer was added to the MCP tool invocation path, requiring every tool call to carry the originating case identifier as a mandatory header. Tool responses were logged with the case ID attached, creating a per-case chain of tool call records that could be traversed forward from case open to recommendation and backward from an adverse action to the specific policy documents retrieved and checks performed. The case ID became the primary audit join key across the orchestrator log, the MCP tool call log, and the adverse-action record.

\textit{Observed outcome}: Following the remediation, internal audit was able to reconstruct the full tool call chain for any case from the case ID alone. The pattern satisfied the ECOA adverse-action audit requirement that the factors considered be traceable and accurately described. The implementation required changes to the MCP tool interface contract, every tool had to accept and propagate the case ID, which meant that tools developed without this constraint needed to be updated before they could be used in a compliant pipeline. This retrofitting cost is noted in Section~\ref{sec:mcpaudit}.

\subsection{Redaction Proxy at the Inference Layer}\label{sec:redaction}

\textit{Problem} (Section~\ref{sec:injection}: Prompt Injection; Section~\ref{sec:residency}: Data Residency): Two distinct threats converge on the model inference endpoint: it is the point at which untrusted document content could carry an injected instruction (Section~\ref{sec:injection}), and it is the hardest data-egress hop to constrain, frequently managed by a third-party service in a different jurisdiction (Section~\ref{sec:residency}). Sending raw customer documents to the model would simultaneously expose the widest injection surface and place resident personally identifiable information (PII) on the least-controllable boundary.

\textit{Mechanism}: Raw documents were never sent to the model. A structured-extraction stage pulled only the specific fields the KYC determination required, and those fields passed through a redaction proxy that removed or tokenized PII before any data reached the inference endpoint. The model reasoned over redacted, structured inputs; the token-to-identity mapping was held within the controlled environment and never transmitted to the model. The proxy and extraction components were sited deliberately within the permitted data-residency boundary, so that the only hop reaching third-party inference infrastructure carried no resident PII and no free-form document text.

\textit{Observed outcome}: The pattern bounded both threats at a single architectural seam. The injection surface was narrowed from arbitrary document bodies to the constrained set of extracted field values (a residual we do not claim was eliminated; see Section~\ref{sec:injection}). The residency obligation was narrowed to the extraction and proxy components, removing the third-party inference endpoint from the scope of resident-PII handling. The cost was a loss of model context: because the model saw only extracted fields rather than full documents, edge cases requiring information outside the predefined extraction schema could not be adjudicated by the agent and were routed to human review, part of the roughly one-in-five referral volume reported in Section~\ref{sec:a2a}.
\section{Formal Model and Analysis of the Control Invariants}\label{sec:formal}

Sections~\ref{sec:threats}--\ref{sec:patterns} describe the control invariants C1--C5 and the architectural patterns through which they are realized. This section reformulates the same concepts in an analytical framework. We (i) give a formal model of a regulated agent execution as a case-bound event chain and express the control invariants C1--C5 as formally testable conditions over that execution model (Section~\ref{sec:model}); (ii) prove that the audit-reconstruction gate of Algorithm~\ref{alg:5} is a necessary and sufficient condition for the invariants to hold, so a compliant outcome is machine-decidable from the case identifier alone (Section~\ref{sec:correctness}); (iii) characterize, for each threat category of Section~\ref{sec:threats}, whether an invariant provably blocks it or only mitigates it, while explicitly delimiting the residual risk (Section~\ref{sec:coverage}); (iv) derive the cost and throughput consequences of the invariants, showing how they shape the performance envelope rather than acting as pure overhead (Section~\ref{sec:cost}); and (v) give an illustrative latency model, with symbolic parameters, whose structure reproduces the bimodal resolution behavior reported in Section~\ref{sec:patterns} (Section~\ref{sec:latency}). This analysis is framework agnostic and applies to other regulated domains (see Section~\ref{sec:generalizing}).

\subsection{Execution Model and Invariants as Predicates}\label{sec:model}

Consider a single regulated agent case, identified by an opaque case identifier ($c$), and comprising all events associated with that case. A regulated execution emits an ordered \emph{event chain} $E(c) = \langle e_1, \ldots, e_n \rangle$, where each event $e_i$ records a consequential step (a tool call, policy retrieval, redaction decision, inference request, or final decision) together with its metadata. Each event carries fields $\mathit{case}(e)$, $\mathit{type}(e)$, $\mathit{policy\_ref}(e)$, and $\mathit{digests}(e)$. We define $A = \{T, P, R, I, D\}$ as the set of event types corresponding to tool, policy, redaction, inference, and decision events, respectively.

The five control invariants are the following predicates over $E(c)$:
\begin{equation}\label{eq:1}
C_1(c) \triangleq \forall e \in E(c):\ \mathit{case}(e) = c
\end{equation}
(\emph{Case-bound continuity}: every consequential step is appended under $c$ and the per-case sequence is gap-free, so no step escapes tagging or ordering.)
\begin{equation}\label{eq:2}
C_2(c) \triangleq \forall e \in E_T(c):\ \mathit{cap}(e) \in \mathit{Allowed}\bigl(\mathit{role}(e), \mathit{wf}(e)\bigr)
\end{equation}
(\emph{Least-privilege}: every tool event uses only a capability allowed for its role and workflow; $E_T$ is the tool-event subset.)
\begin{equation}\label{eq:3}
C_3(c) \triangleq \forall e \in E_P(c):\ \mathit{approved}\bigl(\mathit{policy}_{\mathit{ref}(e)}\bigr) \land \mathit{effective}\bigl(\mathit{policy}_{\mathit{ref}(e)}, t(e)\bigr)
\end{equation}
(\emph{Current-policy grounding}: every policy event references an approved version effective at its timestamp.)
\begin{equation}\label{eq:4}
C_4(c) \triangleq \forall e \in E_I(c):\ \mathit{payload}(e) \in \mathit{Redacted} \land \mathit{pii}\bigl(\mathit{payload}(e)\bigr) = \varnothing
\end{equation}
(\emph{Inference-boundary redaction}: every inference event carries only redacted, PII-free structured payload.)
\begin{equation}\label{eq:5}
C_5(c) \triangleq \mathit{Reconstruct}\bigl(E(c)\bigr) = (\mathit{tools}, \mathit{policy}, \mathit{redaction}, \mathit{factors})
\end{equation}
(\emph{Decision-evidence reconstruction}: the full decision basis is recoverable from the chain.)

\begin{definition}\label{def:1}
A case $c$ is \textbf{compliant} iff $C_1(c) \land C_2(c) \land C_3(c) \land C_4(c) \land C_5(c)$ holds. We write $\Phi(c) \triangleq \bigwedge_i C_i(c)$ for this conjunction.
\end{definition}

\subsection{Correctness of the Audit-Reconstruction Gate}\label{sec:correctness}

Algorithm~\ref{alg:5} returns a status in $\{\mathrm{COMPLETE}, \mathrm{INCOMPLETE}, \mathrm{FAILED}\}$. We show this status determines compliance exactly, provided tool execution is mediated. A deployment is \emph{gate-mediated} if Algorithm~\ref{alg:2} is the \emph{only} path by which a tool executes and every branch of Algorithms~\ref{alg:2}--\ref{alg:4} appends its event to the ledger (as written), each append carrying a per-case sequence number monotone in append order (append-time sequencing). A deployment is additionally \emph{inference-mediated} if every model inference is emitted through the redaction path of Algorithm~\ref{alg:4} and appended as an INFERENCE event carrying its redaction-evidence reference, so that Inference\_Used and each recorded payload faithfully reflect every model call. We additionally assume ledger appends are durable --- an event is persisted before its guarded action proceeds --- and that every regulated workflow issues at least one tool call, so a zero-tool case routes to human review via line~7 of Algorithm~\ref{alg:5}. All\_Control\_Invariants\_Satisfied (Algorithm~\ref{alg:5}, line~9) denotes the computable per-event checks over the chain alone: capability membership in the allowed set for each tool event (C2), digest verification for each event (C1), effective-window containment for each policy reference (C3), and, for each inference event, presence of a redaction-evidence reference together with a PII-free payload (C4). Under append-time sequencing every mediated event is tagged with $c$ by construction, and any omitted or out-of-order event manifests as a gap in the per-case sequence; hence Has\_Sequence\_Gaps (Algorithm~\ref{alg:5}, line~3) decides C1. An event appended under the wrong case constitutes a gate bypass with respect to its true case --- an A1 violation excluded by hypothesis.

\begin{lemma}\label{lem:1}
In a gate-mediated deployment, for every tool event $e \in E_T(c)$ the ledger contains a TOOL\_COMPLETED record iff $e$'s authorization (C2), boundary (C4), and policy-grounding (C3) checks all passed.
\end{lemma}

\begin{proof}
By construction of Algorithm~\ref{alg:2}, a tool is invoked (line~15) only after the three guards at lines~2, 6, and~10 pass; any failure raises before invocation and instead logs TOOL\_AUTHZ\_DENIED, BOUNDARY\_VIOLATION, or POLICY\_GROUNDING\_FAILED. Because the deployment is gate-mediated, no tool executes outside this path, therefore a TOOL\_COMPLETED record exists exactly when all three guards passed. The TOOL\_INVOKED record written at line~14 makes a crash between invocation and completion detectable as an invoked-but-not-completed pair rather than a silent gap. The recorded envelope carries $c$, the capability, the policy reference, and input/output digests, thereby satisfying C1 for that event.
\end{proof}

\begin{theorem}[Soundness and completeness of the gate]\label{the:1}
In a gate-mediated and inference-mediated deployment, Algorithm~\ref{alg:5} returns COMPLETE for case $c$ if and only if $\Phi(c)$ holds; otherwise it returns INCOMPLETE or FAILED, and the case is routed to human review rather than yielding a silent adverse action.
\end{theorem}

\begin{proof}
($\Leftarrow$) Suppose $\Phi(c)$ holds. Then, by append-time sequencing, the per-case sequence is gap-free and every event is well-formed and case-tagged (C1), so lines~2--3 do not return FAILED; the chain contains an approved-and-effective policy event (C3), so line~5 does not return INCOMPLETE; if inference occurred, inference-mediation records an INFERENCE event and C4 guarantees its payload carries a redaction reference and is PII-free, so line~6 does not fire; a decision with a non-empty tool chain exists and, by Lemma~\ref{lem:1}, each tool event is fully recorded, so lines~7--8 pass; and C5 gives reconstructability, so line~9's All\_Control\_Invariants\_Satisfied holds. Algorithm~\ref{alg:5} reaches line~10 and returns COMPLETE. ($\Rightarrow$) We argue by contraposition: if any $C_i$ fails, the corresponding guard in Algorithm~\ref{alg:5} (line~3 for C1, 5 for C3, 6 for C4 (which, under inference-mediation, fires whenever an inference event is missing its redaction record or carries non-redacted PII), 7--8 for the decision/tool chain, 9 for the invariant conjunction) returns FAILED or INCOMPLETE before line~10. Hence $\mathrm{COMPLETE} \Leftrightarrow \Phi(c)$. By Algorithm~\ref{alg:1} lines~17--19 --- and the catch branch at lines~13--14 for raised control exceptions ---, a non-COMPLETE status forces routing to human review, so no outcome is persisted on an unreconstructable basis.
\end{proof}

Theorem~\ref{the:1} provides the analytical counterpart to the argument stated informally in Section~\ref{sec:patterns}: compliance is not merely \emph{documented} but \emph{decidable} from the case identifier, and the failure mode is conservative, routing the case to human review rather than permitting a silent decision. The theorem also delineates the scope of this guarantee: it is conditional on gate-mediation, and a violation of this assumption---namely, a tool-execution path that bypasses Algorithm~\ref{alg:2}---corresponds precisely to the integration hazard identified in Section~\ref{sec:mcpaudit}.

\subsection{Threat Coverage: What the Invariants Block versus Mitigate}\label{sec:coverage}

We distinguish among \textbf{prevention} (the invariant makes the threat's success impossible for a compliant case), \textbf{detection} (the threat may succeed transiently but is guaranteed to surface, under a defined audit-time check, for the subclass that violates policy consistency) and \textbf{mitigation} (the residual is reduced but neither prevented nor guaranteed-detected). Table~\ref{tab:2} summarizes for each threat, identified in Section~\ref{sec:threats}, the strongest assurance established by the invariants and the corresponding residual risk that remains outside their coverage, thereby delineating the scope and limitations of the analysis.

\begin{proposition}\label{pro:1}
Prompt injection surviving redaction (Section~\ref{sec:injection}) is not prevented by C4. Its detectability at audit time is conditional only under a defined check: let $D(E(c))$ compare the factors reconstructed from the decision evidence (C5) against the factors entailed by the applied policy version (C3), and flag $c$ when they diverge. Under gate-mediation, any injection that causes the outcome to deviate from the policy-entailed path triggers $D(E(c))$. This subclass is therefore guaranteed to be detected during audit. Injections that preserve policy consistency produce no such divergence and constitute the residual class not detected by $D$.
\end{proposition}

\begin{proof}
C4 bounds the inference payload to redacted structured fields (Equation~\eqref{eq:4}), however, a crafted value within an allowed field may still reach the model, and prevention is therefore not guaranteed. By Theorem~\ref{the:1}, any persisted outcome has COMPLETE status, so $E(c)$ records both the applied policy version (C3) and the reconstructed decision factors (C5). Consequently, $D(E(c))$ is computable from $c$ alone. 

If an injection causes the outcome to deviate from that entailed by the applied policy, the corresponding factor sets diverge and $D$ flags $c$; this policy-inconsistent subclass is therefore guaranteed to be detected. If, instead, the injected value leads to an outcome that remains consistent with the applied policy, C5 and C3 remain aligned and $D$ does not fire. Such policy-consistent injections constitute false negatives outside the scope of the detection claim. $D$ is an audit-time check external to Algorithms~\ref{alg:1}--\ref{alg:5} and inherits assumption A3 that the referenced policy version faithfully represents the factors entailed by the applicable policy.
\end{proof}

\begin{table}[tbp]
\centering
\caption{Guarantee provided by the control invariants for each threat category. ``Prevent'' = compliant success impossible; ``Detect (conditional)'' = guaranteed to surface under a defined audit-time check, for the policy-inconsistent subclass; ``Mitigate'' = residual reduced only.}\label{tab:2}
\begin{tabularx}{\textwidth}{@{}>{\raggedright\arraybackslash}p{0.22\textwidth}>{\raggedright\arraybackslash}p{0.12\textwidth}>{\raggedright\arraybackslash}p{0.16\textwidth}Y@{}}
\toprule
\textbf{Threat (Section~\ref{sec:threats})} & \textbf{Invariant(s)} & \textbf{Guarantee} & \textbf{Residual not claimed closed} \\
\midrule
A. Prompt injection & C4 + C5 & Detect (conditional) & In-field injected value may reach the model; detected only when it drives the reconstructed factors off the applied policy --- a policy-consistent injection escapes. \\
B. Identity / authorization & C1, C2 & Prevent & Assumes IAM correctly binds roles; mis-provisioned roles are out of model. \\
C. Action auditability & C1, C5 & Prevent & Reconstruction is only as complete as the emitted event schema. \\
D. Tool abuse / over-privilege & C2 & Prevent & Prevents out-of-mandate calls; in-mandate misuse is not a C2 violation. \\
E. Data residency & C4 & Prevent & Bounds the inference hop only; extraction/proxy siting is an assumption. \\
F. Boundary policy enforcement & C2, C3, C5 & Prevent & Holds only under gate-mediation (Theorem~\ref{the:1} precondition). \\
\bottomrule
\end{tabularx}
\end{table}

\subsection{Cost and Throughput Consequences of the Invariants}\label{sec:cost}

The invariants are commonly assumed to be pure compliance overhead. We show that two of them instead \emph{bound} the performance envelope, and the rest add only fixed per-case cost.

\textbf{Audit cost (C1).} Let $E_{\mathit{case}}$ denote the number of events associated with a single case and $E_{\mathit{total}}$ the ledger size. Because C1 makes the case identifier a mandatory join key, reconstruction (Algorithm~\ref{alg:5}) is a case-indexed lookup:
\begin{equation}\label{eq:6}
T_{\mathit{rec}}(c) = O(E_{\mathit{case}}) \ll O(E_{\mathit{total}})
\end{equation}
Since $E_{\mathit{total}}$ grows with deployment age while $E_{\mathit{case}}$ remains bounded for an individual case, the \emph{per-audit} cost of the ECOA reconstruction obligation remains independent of the deployment age rather than scaling linearly with the total ledger size. This property enables the obligation to remain tractable at scale and follows directly from C1 rather than from any specific logging implementation.

\textbf{Throughput (C2).} Let $d_j$ be the per-case call demand on external tool $j$ and $R_j$ its provider-imposed rate limit. Sustainable throughput is bounded by the most restrictive external dependency:
\begin{equation}\label{eq:7}
T \le \min_j \left( \frac{R_j}{d_j} \right)
\end{equation}
Least-privilege (C2) bounds the tool set an agent may invoke and therefore bounds $d_j$; so the invariant that discharges the tool-abuse obligation \emph{also} raises the throughput bound. Over-privileged service accounts (the pre-remediation default of Section~\ref{sec:identity}) inflate the achievable $d_j$ and expose the pipeline to rate-limit exhaustion under buggy or adversarial planning. This coupling is not captured by security treatments that consider least privilege solely as a mechanism for reducing the scope of impact.

\textbf{Marginal overhead (C3, C4).} Redaction adds one extraction-plus-tokenization pass along the in-boundary path; the three gate checks in Algorithm~\ref{alg:2} add $O(1)$ per tool call; each event requires one ledger append. The total per-case compliance overhead is therefore $O(|\mathit{plan}|)$, linear in the (C2-constrained) plan length and independent of ledger size.

\subsection{Illustrative Latency Model (symbolic; not measured)}\label{sec:latency}

To connect the cost analysis to the resolution behavior reported in Section~\ref{sec:patterns}, we introduce an illustrative latency model. Its parameters are symbolic; we introduce no measured values and treat empirical fitting as future work. Under the model, the end-to-end latency of an \emph{automated} case decomposes along Algorithm~\ref{alg:1} as
\begin{equation}\label{eq:8}
L_{\mathit{auto}} = L_{\mathit{init}} + L_{\mathit{red}} + L_{\mathit{pol}} + \sum_{s \in \mathit{plan}} L_{\mathit{tool}}(s) + L_{\mathit{dec}} + L_{\mathit{rec}}
\end{equation}

Here, $L_{\mathit{auto}}$ denotes the total end-to-end latency for an automatically processed case; $L_{\mathit{init}}$ is the latency associated with case initialization and workflow setup; $L_{\mathit{red}}$ is the latency introduced by redaction and boundary-processing operations; $L_{\mathit{pol}}$ is the latency required for policy retrieval and policy-grounding checks; $L_{\mathit{tool}}(s)$ is the latency of external tool invocation $s$; $\mathit{plan}$ denotes the set or sequence of tool invocations executed for the case; $L_{\mathit{dec}}$ is the latency associated with decision generation; and $L_{\mathit{rec}}$ is the latency required to reconstruct and validate the case evidence chain.

Assume (A1) in-boundary operations are $O(1)$--$O(\mathit{fields})$ and small relative to network round-trips, and (A2) external tool calls dominate, so $L_{\mathit{auto}} \approx \sum_s L_{\mathit{tool}}(s)$. By Equation~\eqref{eq:6}, $L_{\mathit{rec}} = O(E_{\mathit{case}})$ is negligible in this sum. Let $p$ be the fraction of cases resolved straight-through and $W$ the human-review SLA. The expected resolution time is
\begin{equation}\label{eq:9}
\mathbb{E}[L] = p \cdot L_{\mathit{auto}} + (1 - p) \cdot (L_{\mathit{auto}} + W)
\end{equation}
Here, $\mathbb{E}[L]$ denotes the expected end-to-end resolution time across all cases. The first term, $p \cdot L_{\mathit{auto}}$, represents cases resolved automatically, while the second term, $(1 - p) \cdot (L_{\mathit{auto}} + W)$, represents cases that first incur the automated-processing latency and are subsequently routed to human review. Thus, $p$ and $1-p$ are the probabilities of automated resolution and human-review routing, respectively. Equivalently, the model defines case-resolution latency as a two-component mixture: a case is resolved in $L_{\mathit{auto}}$ with probability $p$, or in $L_{\mathit{auto}} + W$ with probability $1-p$. Equation~\eqref{eq:9} is the expectation of this mixture. Under the simplifying assumption that within-path latency variation is small relative to the separation introduced by human review, the resulting distribution is bimodal, with a fast component near $L_{\mathit{auto}}$ and a slow component near $L_{\mathit{auto}} + W$, having respective masses $p$ and $1-p$.

Two consequences follow directly. First, reducing the review fraction $1-p$ (e.g., by widening the extraction schema described in Section~\ref{sec:redaction}) reduces the mass of the slow mode rather than its latency. Consequently, mean resolution time is governed primarily by routing behavior rather than by per-case latency reductions. Second, by Equation~\eqref{eq:7}, the sustained case rate is bounded by $\min_j(R_j/d_j)$. Thus, system scale is constrained by the tightest external dependency and by C2's bound on $d_j$, rather than by human-review headcount, consistent with the horizontal-scaling behavior reported in Section~\ref{sec:a2a}.

\subsection{Scope and Limitations of the Compliance Guarantee}\label{sec:scope}

The results presented in earlier sections are conditional. We state the assumptions required by each result, the approximations introduced by the latency model, and the conditions under which Theorem~\ref{the:1} no longer holds. This delimits the guarantee and identifies the residual attack surface a deployer must close by other means.

\textbf{Assumptions of Theorem~\ref{the:1}. A1 --- Gate-mediation.} Under the gate-mediation assumption, every tool invocation is required to traverse the control path defined by Algorithm~\ref{alg:2}. If a side channel invokes a tool without traversing the gate through a direct API call, an un-instrumented SDK, or an out-of-band script, the ledger does not observe the execution, Lemma~\ref{lem:1} no longer holds, and Algorithm~\ref{alg:5} may return COMPLETE for a case whose execution violates an invariant. This is the integration hazard reported in Section~\ref{sec:mcpaudit} and represents the dominant practical failure mode. The same mediation requirement applies to inference: every model call must traverse the redaction path of Algorithm~\ref{alg:4} and emit an INFERENCE event, ensuring that an un-instrumented inference cannot leave a C4 violation unobserved.

\textbf{A2 --- Schema completeness.} The event schema records every field evaluated by the predicates, including capability, policy reference, boundary class, and digests. If an event omits a tested field, $\mathit{Reconstruct}$ is under-defined and C5 may be satisfied vacuously; the guarantee then degrades to being only as complete as the schema.

\textbf{A3 --- Oracle correctness.} The predicate oracles --- \emph{Authorize}, \emph{Boundary\_Enforcer.Allow}, the \emph{approved}/\emph{effective} policy state, and the IAM role binding, are assumed correct. A mis-provisioned role (C2), a mislabelled policy state (C3), or a false negative from the redaction classifier (C4) causes a guard to pass when it should fail. Theorem~\ref{the:1} certifies \emph{consistency with the oracles}, not the correctness of the oracles themselves.

\textbf{A4 --- Boundary siting.} The extraction and redaction-proxy components are assumed to reside inside the permitted data-residency boundary (Section~\ref{sec:residency}). If those components are themselves hosted across the boundary, C4 holds at the inference hop while residency is nonetheless violated upstream.

\textbf{A5 --- Ledger integrity.} The audit ledger is append-only and tamper-evident. If an adversary can delete or rewrite events, \emph{Fetch\_By\_Case} returns a chain that no longer reflects execution and the equivalence in Theorem~\ref{the:1} breaks. This is a standard secure-logging assumption and is orthogonal to, rather than established by, the present construction.

\textbf{Approximations in the latency model.} Equation~\eqref{eq:8} is exact as a decomposition, but the reduction to a tool-latency sum assumes in-boundary operations are negligible against network round-trips (A1--A2 of Section~\ref{sec:latency}). Where extraction or redaction is expensive, for example for large documents or heavy tokenization, the approximation underestimates end-to-end latency. Equation~\eqref{eq:9} treats the straight-through fraction $p$ and the review service interval $W$ as constants and assumes cases to be independent. It is therefore a first-moment model that does not capture queueing dynamics, contention in the review queue, correlations between case difficulty and $W$, or any dependence of $p$ on system load. Equation~\eqref{eq:7} similarly treats provider rate limits as fixed and independent; it therefore represents an upper bound under static limits rather than a predictive model of realized throughput.

\textbf{Where the theorem fails.} Theorem~\ref{the:1} fails, and Algorithm~\ref{alg:5} may certify a non-compliant case as COMPLETE, precisely when any of A1--A5 is violated: a tool path bypasses the gate (A1); the event schema omits a tested field (A2); an oracle is incorrect (A3); a proxy is sited across the residency boundary (A4); or the ledger is mutable (A5). Two additional limitations are inherent to the model rather than violations of its assumptions. First, the guarantee establishes reconstructability, not substantive correctness: a fully reconstructable chain may still record an incorrect KYC decision. Second, the model does not capture in-mandate misuse: a capability exercised within its authorized set but for an inappropriate purpose does not constitute a C2 violation. Addressing these limitations lies outside the scope of the present model and motivates the mechanized-verification direction proposed in \cite{rashie2026typechecked}.

\subsection{Summary}\label{sec:summary}

The invariants of Section~\ref{sec:threats} are not only descriptive but (i) decidable --- Theorem~\ref{the:1} makes a compliant outcome a verifiable function of the case identifier; (ii) explicitly delimited --- Table~\ref{tab:2} separates prevention from detection and names each residual; and (iii) performance-shaping --- C1 bounds audit cost (Equation~\eqref{eq:6}) and C2 bounds throughput (Equation~\eqref{eq:7}). We regard Theorem~\ref{the:1} and the C1/C2 cost bounds as the transferable core: any regulated agent deployment that enforces case-bound, gate-mediated execution inherits the same decidable-audit and cost-envelope properties, independent of the domain.
\section{Negative Results and Open Problems}\label{sec:negative}
Systematic reporting of unsuccessful outcomes is among the most valuable contributions a practitioner systematization can make, because it is precisely what laboratory studies cannot observe. This section records three results from the deployment described in Section~\ref{sec:patterns}: a control failure caught by internal audit, an architectural gap whose remediation carried an ongoing cost, and a fundamental design limit that no mitigation resolved. Table~\ref{tab:3} summarizes the three results.

\subsection{Stale Policy in the RAG Store Caused Systematic Over-Verification}\label{sec:stalepolicy}

The most consequential failure in the deployment was not a security attack but a policy synchronization gap. The compliance team updated the KYC policy to remove the step-up verification requirement for applicants above a specified credit score threshold. The automated compliance feed delivered the updated document, but the human sign-off gate described in Section~\ref{sec:rag} had not yet approved the new version for production ingestion. During the window between the policy update and the gate approval, the agent continued retrieving the prior policy version and applying step-up KYC to customers who no longer required it under current policy.

The failure was identified by internal audit, rather than by the agent system or its operational monitoring. The audit finding characterized it as a control gap: the agent was acting on a superseded policy provision, producing outcomes that did not accurately reflect the institution's current compliance posture. The regulatory exposure was in the direction of over-verification rather than under-verification---affected customers were subjected to additional friction that the updated policy did not require---which raised a potential Unfair, Deceptive, or Abusive Acts or Practices (UDAAP) concern independent of any security incident.

The broader implication is that a human sign-off gate can protect against the ingestion of invalid or unapproved policy content but does not, by itself, eliminate the synchronization interval preceding approval. In a regulated deployment, this interval constitutes a compliance gap rather than merely an operational delay. Current agent frameworks do not natively provide mechanisms for synchronizing policy versions with the compliance-document lifecycle, leaving this as an open systems problem.

Expressed in terms of C3, the failure is that \emph{Select\_Highest\_Version} (Algorithm~\ref{alg:3}) ranged over \emph{approved} documents while the compliance team had already retired the encoded provision. A minimal strengthening is to make effective dating authoritative over approval recency: candidate policies are first filtered according to the compliance-authored effective window, while the presence of a delivered but not yet approved successor is treated as evidence that the incumbent is pending supersession. Affected cases are then routed to human review (INCOMPLETE) rather than evaluated against the stale policy version. This modification preserves the sign-off gate while preventing approval latency from silently creating a period of policy inconsistency. The synchronization mechanism directly addresses the open problem identified above.

\begin{table}[tbp]
\centering
\caption{Negative results from the deployment: what failed, how it surfaced, its root cause, and the open problem or generalizable lesson. Two of the three were surfaced by internal audit or production observation rather than by adversarial testing or operational monitoring: the class of failure laboratory studies cannot observe.}\label{tab:3}
\begin{tabularx}{\textwidth}{@{}>{\raggedright\arraybackslash}p{0.17\textwidth}>{\raggedright\arraybackslash}p{0.15\textwidth}YY>{\raggedright\arraybackslash}p{0.07\textwidth}@{}}
\toprule
\textbf{Failure} & \textbf{How It Surfaced} & \textbf{Root Cause} & \textbf{Open Problem / Lesson} & \textbf{Section} \\
\midrule
A. Stale policy in the RAG store caused systematic over-verification & Internal audit (not the agent or its monitoring) & Policy update delivered by the automated feed, but the human sign-off gate had not yet approved the new version; the agent kept serving the prior one. & A sign-off gate protects against ingesting bad content but not the window before it closes. No framework natively synchronizes policy versions with a compliance-document lifecycle. & \ref{sec:stalepolicy} \\
\addlinespace
B. MCP tool interfaces were not designed for audit & Production integration of the Case-ID propagation layer (Section~\ref{sec:caseid}) & Tools built without audit scope carry no case/session/provenance in their signatures, forcing a retrofit of every interface contract. & Auditability in a regulated MCP deployment is an interface-design constraint that must precede tool construction, not an infrastructure-layer logging concern. & \ref{sec:mcpaudit} \\
\addlinespace
C. $\approx$1 in 9 legitimate applicants could not be served & Production observation across applicant population & Mandatory two-channel 2FA; single-channel applicants have no path; a lost-credential exception path exceeded product economics. & No regulatory framework defines what an institution owes applicants excluded by an automated system's authentication preconditions; a disparate-impact question remains unresolved. & \ref{sec:oneinnine} \\
\bottomrule
\end{tabularx}
\end{table}

\subsection{MCP Tool Interfaces Were Not Designed for Audit}\label{sec:mcpaudit}

The case-ID propagation layer described in Section~\ref{sec:caseid} resolved the audit trail gap, but its implementation revealed a structural problem in how MCP tools are specified and developed. Tools built without audit requirements in scope do not carry case identifiers, session context, or any provenance information in their call signatures. Adding the case-ID constraint to the production pipeline required retrofitting every existing tool to accept and propagate the identifier, a substantial change to the interface contract that forced re-testing and re-validation of each tool before it could be used in a compliant workflow.

The broader implication is that auditability in a regulated MCP deployment is not a logging concern that can be addressed at the infrastructure layer; it is an interface design constraint that must be established before tools are built. The academic literature on MCP security does not address this: the systematizations surveyed in Section~\ref{sec:background} treat tool calls as atomic units to be authenticated and authorized, without considering the provenance and traceability requirements that regulated deployments impose on the tool call record. Organizations adopting MCP in regulated contexts will encounter this gap at the integration stage, and not at the design stage, which imposes additional cost.

\subsection{Roughly One in Nine Applicants Could Not Be Served by the Automated Pipeline}\label{sec:oneinnine}

The deployment assumed that applicants had at least two active contact channels (a mobile number and an email address) because two-factor authentication was effectively mandatory for the consumer credit product. Roughly one in nine applicants presented with only one active channel, typically because they had lost access to or could not recall a previously registered email address. These applicants could not complete the automated KYC pipeline regardless of their creditworthiness or identity verification status. This one-in-nine group is distinct from the approximately one-in-five admitted cases that the automated path routes to human review (Section~\ref{sec:a2a}): the former are excluded at intake because a second authentication channel is unavailable and therefore never enter the processing pipeline, whereas the latter are admitted cases that enter the pipeline and are subsequently routed to review.

The pattern of accommodating single-channel applicants was evaluated and abandoned. The security posture required for a consumer credit product did not permit single-channel 2FA, and the operational cost of building a separate exception path for lost-credential recovery exceeded what the product economics supported. The result is a real population of legitimate applicants, roughly one in nine, for whom the automated system provides no path to approval. This is not a failure of the agent architecture; rather, it is a design limitation that the architecture itself cannot resolve and that existing academic threat models do not capture, because laboratory evaluations generally do not model users who are unable to satisfy the required authentication preconditions.

The open problem this exposes is the absence of a regulatory framework for what an institution owes applicants who are excluded by the authentication requirements of an automated system. 

ECOA prohibits discrimination on specified bases; it does not directly address systemic exclusion arising from technical preconditions. Whether a dropout rate of approximately one in nine attributable to 2FA channel requirements constitutes a disparate-impact concern remains unresolved under the evolving regulatory framework: the 2026 amendments to Regulation B remove the effects test, stating the Bureau's position that ECOA does not authorize disparate-impact liability, while noting that creditors remain subject to state laws analogous to ECOA \cite{cfpb2026regb}. What obligations an institution has toward applicants excluded by such a technical precondition remains a question that neither the security literature nor current regulatory guidance has resolved.
\section{Generalizing Beyond Finance}\label{sec:generalizing}
The control obligations that dominate regulated finance (auditability of automated decisions, data residency, least-privilege authorization, and explainability for high-stakes outcomes) are not unique to finance. They recur, with sector-specific variation, across healthcare (HIPAA), critical infrastructure, and the EU AI Act's high-risk categories \cite{tabassi2023airmf,iso42001}. The mapping developed in Section~\ref{sec:threats} and the patterns of Section~\ref{sec:patterns} are therefore expected to generalize: the underlying threat categories are largely domain-independent, whereas the degree of regulatory amplification and the acceptable cost frontier vary by sector. We conjecture that auditability and least-privilege tool authorization are the two controls whose production implementation is both challenging and broadly generalizable, as both require mechanisms (immutable per-action evidence trails and enforced capability bounds) that the current agent frameworks do not provide natively \cite{chu2026survey,gaire2025sok}. A systematic cross-domain validation of this conjecture is a focus of our future work.
\section{Conclusions}\label{sec:conclusions}
Agentic AI is entering regulated financial systems faster than the security literature has connected its threat models to the binding obligations of this domain. We have argued that the gap between laboratory taxonomy and regulated production is itself the problem worth addressing and have offered a novel and unique experience-grounded bridge: a mapping from established agentic threats to regulatory control obligations, a set of architectural patterns observed in a production regulated-finance deployment, and a complete account of their limitations. The central implication is that securing agents under regulation is less about novel attack classes than about making auditability, least-privilege authorization, and boundary policy enforcement real at production scale: requirements that the current agent frameworks leave to the deploying organization. This work is intended to support further peer-reviewed, deployment-grounded research at the intersection of agent security and regulatory practice.
\section*{Disclosure}

We describe only generally applicable architectural patterns and publicly mandated regulatory obligations, and disclose no proprietary, confidential, or customer-identifying information of any organization; the views are the authors' own.

\section*{Conflict of interest}

The authors declare no competing financial or non-financial interests.

\section*{Funding}

This research received no specific grant from any funding agency in the public, commercial, or not-for-profit sectors.

\section*{Data availability statement}

The reference implementation of the control invariants and algorithms, together with its synthetic dataset and test suite, is openly available at \artifacturl{} \cite{casebound_artifact}; Appendix~\ref{app:artifact} describes its contents. The production observations reported in this paper are not derived from that artifact. The underlying operational and internal-audit records of the production deployment are confidential to the deploying institution and are not available for release.

\section*{AI usage statement}

The authors used generative AI tools (Claude Opus 4.8, Claude Opus 5 and Claude Opus 5.5; Anthropic, anthropic.com) for language editing and drafting assistance of the manuscript, for generating the synthetic dataset, and to assist in developing the reference implementation's software. The authors reviewed all AI-assisted output and take full responsibility for the content of this article.

\section*{Ethics statement}

This work is a retrospective practitioner account of a production system, not human-subjects research, and did not require institutional approval. No personal, customer-identifying, or proprietary data is accessed for, or contained in, this paper; only aggregate, de-identified operational observations are reported, consistent with the inference-boundary redaction controls described in Section~\ref{sec:redaction}.

\bibliographystyle{unsrtnat}
\bibliography{references}

\appendix

\section{Reference Implementation and Artifact}\label{app:artifact}

The control invariants of Section~\ref{sec:invariants}, the reference algorithms of Section~\ref{sec:algorithms} and the gate of Theorem~\ref{the:1} are implemented in an open-source reference implementation, \emph{casebound}, available at \artifacturl{}. This appendix describes what the repository contains, how its parts correspond to the paper, and how to reproduce its checks. The implementation runs on synthetic data only. It is not the production system described in Sections~\ref{sec:patterns} and~\ref{sec:negative}, and none of the production observations reported there were produced by it.

\subsection{Repository contents}

The repository holds three Python packages (Table~\ref{tab:artifact-packages}). The core package depends only on the Python standard library. Everything that needs a third-party framework (model provider SDKs, the Model Context Protocol) lives in a separate package and reaches the core only through the interfaces the core defines, so the algorithms can be read and tested without any external service.

\begin{table}[h]
\centering
\caption{Packages in the artifact repository.}\label{tab:artifact-packages}
\begin{tabularx}{\textwidth}{@{}>{\raggedright\arraybackslash}p{0.2\textwidth}Y@{}}
\toprule
\textbf{Package} & \textbf{Contents} \\
\midrule
\texttt{casebound} & The core: event schema and hash-chained ledger, Algorithms~\ref{alg:1}--\ref{alg:5}, the predicate $\Phi(c)$, and a gated model adapter that is the only path to a model. A synthetic know-your-customer (KYC) workload, a command-line demonstration, a single-file browser demonstration, and the test suite. \\
\texttt{casebound-models} & Model clients behind the core's model interface: Claude through the Anthropic API or Google Vertex AI, and Gemini through Google Vertex AI. A record-and-replay facility lets a recorded model run be replayed without network access or credentials. \\
\texttt{casebound-mcp} & The tool-call contract of Section~\ref{sec:caseid} for the Model Context Protocol (MCP): tool servers that refuse a call without a case identifier, a client adapter, and a gateway that applies Algorithm~\ref{alg:2} to every call from an MCP client. \\
\bottomrule
\end{tabularx}
\end{table}

\subsection{Correspondence with the paper}

Table~\ref{tab:artifact-map} maps the paper's constructs to the code that implements them and the tests that check them.

\begin{table}[h]
\centering
\caption{Where each construct of the paper is implemented and checked. Module paths are relative to \texttt{packages/casebound/src/casebound/} unless stated otherwise.}\label{tab:artifact-map}
\begin{tabularx}{\textwidth}{@{}>{\raggedright\arraybackslash}p{0.3\textwidth}>{\raggedright\arraybackslash}p{0.26\textwidth}Y@{}}
\toprule
\textbf{Paper} & \textbf{Implementation} & \textbf{Checked by} \\
\midrule
Algorithm~\ref{alg:1}, Execute\_Regulated\_Workflow & \texttt{workflow.py} & Scenario tests \\
Algorithm~\ref{alg:2}, Invoke\_Controlled\_Tool (C2) & \texttt{gateway.py} & Scenario and tool-boundary tests \\
Algorithm~\ref{alg:3}, Retrieve\_Current\_Policy (C3) & \texttt{policy.py} & Scenario tests \\
Algorithm~\ref{alg:4}, Redact\_Before\_Inference (C4) & \texttt{redaction.py}, \texttt{model.py} & Payload-scan and model-gate tests \\
Algorithm~\ref{alg:5}, Reconstruct\_Decision\_Evidence (C5) & \texttt{reconstruct.py} & Scenario and paper-alignment tests \\
Hash-chained, case-bound ledger (C1) & \texttt{ledger.py}, \texttt{events.py} & Tampering scenario \\
$\Phi(c)$ and Theorem~\ref{the:1} & \texttt{reconstruct.py} (\texttt{phi}) & Property-based test over randomly generated executions \\
Case-ID propagation (Section~\ref{sec:caseid}) & \texttt{casebound-mcp} package & Contract, gateway and transport tests \\
Inference mediation (Assumption A1) & \texttt{model.py} & Model-gate tests \\
\bottomrule
\end{tabularx}
\end{table}

The property-based test generates random gate-mediated executions, varying the allowed capability set, the plan, and the policy's approval state and effective window, and asserts that Algorithm~\ref{alg:5} returns COMPLETE exactly when $\Phi(c)$ holds.

\subsection{Fault scenarios}

The demonstration workload defines one scenario per guarantee or failure mode the paper discusses. Each one runs the full workflow and then reconstructs the case from its identifier alone. The scenarios cover a compliant case; a stale policy served from the retrieval index (Section~\ref{sec:stalepolicy}) and the effective-dating remedy; a tool called outside Algorithm~\ref{alg:2} (Section~\ref{sec:mcpaudit} and Assumption A1); an inference payload that skipped redaction or still carried personal data after it; a plan that requests a capability outside the agent's role; a capability revoked after the case closed; a worker failure between tool invocation and completion; a ledger event altered after the fact; an instruction embedded in an application field that steers the decision (Proposition~\ref{pro:1}); and a plan with no tool calls. The expected verdict of every scenario is fixed in the test suite.

\subsection{Synthetic data}

All data in the repository is synthetic. A seeded generator produces a population of applicants and a corpus of versioned policies, so the dataset is reproducible and can be reviewed as a diff. Its parameters (score distribution, mismatch rates, policy thresholds and dates) are illustrative: the outcome shares the demonstration computes are not evidence for the production observations reported in this paper. As in Section~\ref{sec:oneinnine}, applicants with a single contact channel are excluded at intake and never become cases. Generated identifiers are drawn from ranges reserved for fiction (for example, Social Security numbers in the 900--999 area, which is never issued, telephone numbers in the 555-01XX range and e-mail addresses under the reserved \texttt{.test} domain). The tool back ends are deterministic stand-ins that answer from the generated data.

\subsection{Reproducing the checks}

The packages require Python~3.11 or later. One test runs the browser demonstration's engine in Node.js to confirm it agrees with the Python core, and it is skipped when Node.js is not installed. No credentials or network access are needed: the model-client tests use a recorded run, and calls to live models are opt-in. Each command below runs from the repository root, in a fresh virtual environment and a POSIX shell:

\begin{verbatim}
pip install -e "packages/casebound[dev]"
(cd packages/casebound && pytest -q)   # incl. Theorem 1
casebound-demo --list          # scenarios and expected verdicts
casebound-demo --scenario stale-policy
\end{verbatim}

The model-client and MCP packages have their own suites:

\begin{verbatim}
pip install -e "packages/casebound-models[dev,vertex]"
pip install -e "packages/casebound-mcp[dev]"
(cd packages/casebound-models && PYTHONPATH=../casebound pytest -q)
(cd packages/casebound-mcp && PYTHONPATH=../casebound pytest -q)
casebound-demo --tools mcp --scenario happy
\end{verbatim}

The repository's continuous-integration configuration runs the same commands on every change.

\end{document}